\documentclass[pra,onecolumn,floatfix,a4paper,superscriptaddress]{revtex4}
\usepackage{bm,color,graphicx,amsmath,txfonts}

\usepackage[colorlinks, citecolor=blue,linkcolor=blue]{hyperref}

\newcommand{\e}{{e}}

\begin{document}
\title{Enhanced Gaussian interferometric power, entanglement and Gaussian quantum steering in magnonics system with squeezed light }

\author{Noura Chabar}
\thanks{chabarnoura@gmail.com}
\affiliation{LPTHE-Department of Physics, Faculty of sciences, Ibnou Zohr University, Agadir, Morocco}	
\author{M'bark Amghar}
\affiliation{LPTHE-Department of Physics, Faculty of sciences, Ibnou Zohr University, Agadir, Morocco}	
\author{Mohamed Amazioug} 
\thanks{amazioug@gmail.com}
\affiliation{LPTHE-Department of Physics, Faculty of sciences, Ibnou Zohr University, Agadir, Morocco}

\begin{abstract}

In this study, we propose a scheme to improve quantum correlations (QCs) between two magnons in a tripartite magnonical system.  We use Gaussian interferometric power (GIP) to quantify QCs beyond entanglement. Additionally, Gaussian quantum steering is discussed. We investigate the enhancement of QCs via a squeezing parameter and an optical parametric amplifier (OPA). The Mancini criterion is considered to confirm the presence of shared entanglement between the two magnons. We observed a squeezing of about 7dB for the first  magnon. Additionally, the squeezed vacuum field and the OPA improve the generation of genuine tripartite entanglement. We hope that current experimental technology will allow the proposed scheme to be implemented.\\


	\end{abstract}
	
	\date{\today}
	
	\maketitle
	
\section{Introduction} 
 In recent years, Yttrium iron garnet (YIG), an exceptional ferrimagnetic component, has generated substantial interest. The magnons are quanta of collective spin excitations in materials like  YIG  \cite{yig}. High spin density in YIG makes it possible for microwave cavity photons and magnons to interact strongly, creating cavity-magnon polaritons \cite{2,4,5,6,7,8}. The Kittel mode \cite{1kiteel} in YIG has unique properties, including significant magnonic nonlinearities \cite{2} and an incredibly low damping rate \cite{2}. Significant coupling between the YIG sphere and cavity photons has been reported at cryogenic and ambient temperatures \cite{8}. All of these improved qualities will be supported by the forthcoming quantum information networks \cite{9}. So, many interesting things have been studied in the field of magnons, such as bistability \cite{10}, the attraction of energy levels in cavity magnon-polaritons \cite{12}, progress in cavity spintronics \cite{7,11}, the appearance of magnon dark modes \cite{13}, the intriguing idea of the exceptional point \cite{14}, and more. Magnons have a strong coupling, which has led to the investigation of several elements of quantum information, such as the coupling of the superconducting qubit to magnons \cite{15} and phonons \cite{16}. Other fascinating phenomena include the transparency induced by magnons \cite{17}, magnetically controllable slow light \cite{18}, and  magnon blockade \cite{annal}.\\ \\
Lately, researchers have been studying the generation of spin currents in one YIG sample as a result of excitation occurring in another YIG sample \cite{11}. This stems from the coupling between the two samples, facilitated by the cavity mediation \cite{lien}. It is obvious to consider the potential for quantum entanglement between two YIG samples, as the study of quantum entanglement between macroscopic systems has attracted a great deal of interest  \cite{ Ann, lien}. The detection of quantum entanglement between macroscopic mechanical oscillators has recently proved remarkably successful \cite{22,23} using photonic crystal cavities and with superconducting qubits \cite{super1, super}. Furthermore, there have been documented instances of entanglement between the cavity field and mechanical motion \cite{24}. The widely used method for creating entanglement incorporates nonlinearities inside the system \cite{ secondnon}. Nonlinearities that are widely recognized include the magnetostrictive interaction \cite{16} and the Kerr effect \cite{2}. The magnetostrictive force facilitates the coupling of magnons to phonons, offering a means to generate entanglement between magnons and phonons \cite{20}. The  Kerr nonlinearity originates from magnetocrystalline anisotropy and has been effectively employed to induce bistability in magnon-photon systems. In recent work, these nonlinearities have been utilized to induce entanglement between two magnon modes within a magnon-cavity system \cite{25,28,29}.\\ \\
The several theoretical studies and how to implement QCs between two magnons constitute a motivating database that can be utilized to enhance these correlations in microwave cavities \cite{25dens, onewayster}. In the proposed system, we can generate genuine tripartite entanglement even in the absence of magnon deformation(phonon). Furthermore, as indicated in Ref \cite{25dens}, the absence of magnon-phonon coupling also results in the absence of magnon-magnon entanglement. Nevertheless, our approach has effectively established entanglement between two magnons in the absence of magnon-phonon coupling.  
 \\ \\ 
In this work, we propose a system to improve QCs between two magnons in a  cavity magnonic. Our method takes advantage of the nonlinearity of the optical parametric amplifier and involves driving the cavity with a squeezed vacuum field. The couple of YIG spheres are connected to the cavity field through the magnetic dipole interaction  and the cavity is driven by a squeezed vacuum field. The squeezing vacuum noise within the cavity leads to the generation of squeezed states of magnons. Squeezed states serve not only to enhance measurement sensitivity \cite{squerole1} and investigate decoherence theories on a large scale \cite{squerole2}, but also constitute a fundamental component in continuous variable information processing \cite{squezim}. The robustness of entanglement against thermal effects is significantly improved by magnon squeezing \cite{fb, sohail}. We have demonstrated that squeezed magnon states can be achieved. Also, a genuine tripartite entanglement state involving the two magnons and photons is possible in the presence of both the squeezed vacuum field and intracavity squeezed light, within experimentally accessible parameters. We show the robustness of the entanglement under the effect of the thermal noise by adjusting both the non-linear gain and the value of the squeezing parameter $r$. The bipartite magnon-magnon entanglement is quantified using logarithmic negativity. In order to broaden our investigation outside of entanglement, we employ GIP  and  Gaussian quantum steering. We investigate the behavior of quantum steering in the steady state regime, which is the asymmetric characteristic observed between two entangled observers, enabling one participant (Alice) to manipulate or "steer" the state of a remote participant (Bob) by exploiting their exchanged entanglement \cite{steer}. Our findings indicate that the squeezing parameter and non-linear gain are crucial to manipulating genuine tripartite entanglement.\\

The paper's structure is organized in the following manner: Section II outlines our system's model and Hamiltonian. Then, we express the Hamiltonian, and we give the Langevin quantum equations (QLEs) describe the system. Section III presents the quantifiers used to measure the different correlations between the magnon modes, as well as the function used to quantify tripartite entanglement. Section IV is focused on examining the squeezed state of one magnon mode, the Mancini criteria for proving the presence of a sharing entanglement between the two magnons, and the quantifiers used to measure the different correlations between the magnon modes. Finally, Section V concludes our findings.

\section{Model and equations}

\begin{figure}[h]

	\includegraphics[width=0.62\linewidth]{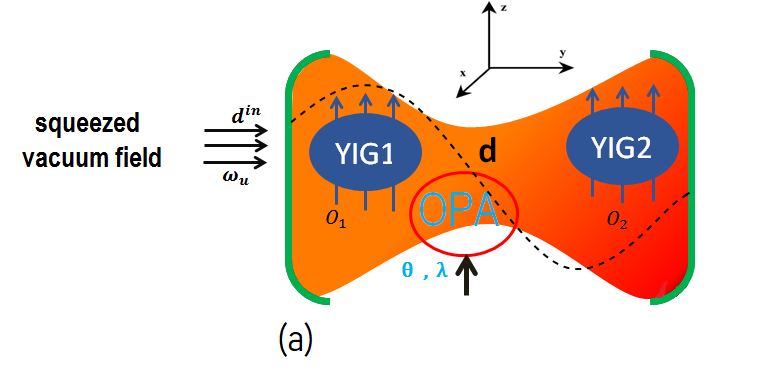}
	\hfil
\includegraphics[width=0.37\linewidth]{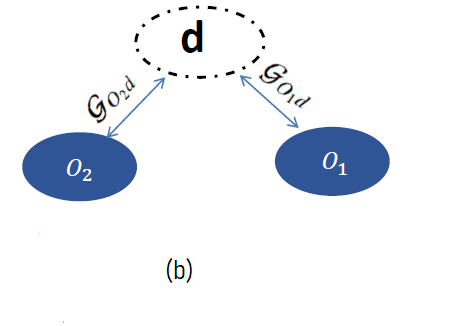}
\caption{(a) A schematic diagram of two YIG spheres placed inside a microwave cavity situated in proximity to the maximum magnetic field of the cavity mode and concurrently exposed to a uniform bias magnetic field. The cavity is driven by a squeezed vacuum field. The magnetic field of the cavity mode is in the x direction, and the bias magnetic field is applied along the z direction. Inside the microwave cavity, we inject an optical parametric amplifier. The non-linear gain and phase of the OPA are denoted by $\lambda$ and $\theta$, respectively. The input is at the frequency $\omega_u$. The diagram (b) shows the couplings within the system.  $\mathcal{G} _{O_{1}d}$ and $\mathcal{G} _{O_{2}d}$ are the linear coupling forces between photons and magnons. The cavity is driven by a squeezed vacuum field generated by a Josephson parametric amplifier, which doesn't show up in the figure.} 

	\label{fig:syst-hp2} 
\end{figure}

The system under consideration is composed of cavity microwave photons and magnons, as illustrated in Fig. (\ref{fig:syst-hp2}). The cavity is driven by a squeezed vacuum field. YIG spheres are characterized by extended lifetime, elevated spin density, low damping \cite{lifetimelongmagnon}, etc. In the proposed system, the excessive spin density in YIG spheres enables strong coupling between magnons and photons via the magnetic dipole interaction. We assume that the size of the two YIG spheres is significantly smaller than the microwave field wavelength, thus rendering the impact of radiation pressure negligible \cite{16}. 

The system's Hamiltonian is expressed as follows (with $\hbar=1$)       	
\begin{equation}
		\begin{aligned}
			\hat{\mathcal{H}}  & =\omega_d  \hat {{d}}^{\dagger} \hat d+\omega_{O_1} \hat O_1^{\dagger} \hat O_1+\omega_{O_2} \hat O_2^{\dagger} \hat O_2 
			 +\mathcal{G}_{O_1 d}\left(\hat d+ \hat d^{\dagger}\right)\left(\hat O_1+ \hat O_1^{\dagger}\right)+\mathcal{G}_{O_2 d}\left(\hat d+\hat d^{\dagger}\right)\left( \hat O_2+ \hat O_2^{\dagger}\right)  + i\lambda \left( \hat d^{{\dagger}{2}} \e^{-i 2 \omega_{u}t} - \hat d^{2}\e^{i 2\omega_{u}t}\right),
		\end{aligned}
\end{equation}
the operators $ \hat{d}$ and $ \hat{d}^{\dagger}$ correspond in this order to the annihilation and creation operators of cavity mode; the operators $\hat{O}_{1},\hat{O}_{2}$ $( \hat{O}_{1}^{\dagger}, \hat{O}_{2}^{\dagger})$ represent, respectively, the annihilation (creation) operators of the two magnon modes. These magnon modes represent the collective movement of spins achieved via the Holstein-Primakoff transformation \cite{1}. The parameter $ \omega_{d}$ and $\omega_{ O_{p}}$ $(p=1,2)$ are respectively the resonance frequencies of the cavity and the magnon modes. The formula for the magnon frequency is given by the following expression: $\omega_{O_{p}}= \gamma H_{p}$, the parameters $\gamma/\pi= 28$ GHz/T represent the gyromagnetic ratio, and $H_{p}$ are the external bias magnetic fields. The linear coupling forces between photons and magnons are represented by the term $\mathcal{G}_{O_{p}d}$.  The behavior of the system's dynamics can be articulated using the quantum Langevin formalism. Adopting the rotating wave approximation the term $\mathcal{G}_{O_p d}\left(\hat{d}+\hat{d}^{\dagger}\right)\left(\hat{O}_p+ \hat{O}_p^{\dagger}\right)$ becomes $\mathcal{G}_{O_p d}\left(\hat{d} ~ \hat O^{\dagger}_{p}+ \hat{d}^{\dagger} \hat{O}_{p}\right)$ (valid when $ \omega_{O_p}, \omega_d \gg \mathcal{G}_{O_{p} d}, k_d, k_{O_{p}} $) as mentioned in \cite{20}, with $\Delta_d=\Delta_{O_p}$. 
In the rotating frame at the frequency $\omega_u$ of the squeezed vacuum field, the QLEs governing the system can be formulated as follows 

\begin{equation}
		\begin{aligned}
		\dot{\hat d}  =-\left(i \Delta_d+ k_d\right) \hat d-i \mathcal{G}_{O_1 d} \hat O_1-i \mathcal{G}_{O_2 d} \hat O_2+2 \lambda \hat d^{\dagger} +\sqrt{2 k_d} \hat d^{i n},
		\end{aligned}
\end{equation}
\begin{equation}		
		\begin{aligned}
\dot{\hat O}_1  =-\left(i \Delta_{O_1}+k_{O_1}\right) \hat O_1-i \mathcal{G}_{O_1 d} \hat d+\sqrt{2 k_{O_1}} \hat O_1^{i n},
	\end{aligned}
\end{equation}
\begin{equation}		
	\begin{aligned}
\dot{\hat O}_2  =-\left(i \Delta_{O_2}+k_{O_2}\right) \hat O_2-i \mathcal{G}_{O_2 d} \hat d+\sqrt{2 k_{O_2}} \hat O_2^{i n}
\end{aligned}.
\end{equation}
where the detunings $\Delta_{d}=\omega _{ d}-\omega_u $ and $\Delta_{O_p}=\omega _{ O_p}-\omega_u $, $k_{d}$ is the dissipation rate of the cavity and $k_{O_{p}}$ ($k_{O_{1}}$ and  $k_{O_{2}}$ ) are the dissipation rates of magnon modes . The parameters $ \hat d^{i n}$ and $ \hat O_{p}^{i n}$ are the input noise operators of the cavity and magnon modes, respectively. The input noise operators have a zero mean and are characterized by the following  correlation properties \cite{noise correlations}: $\left\langle \hat d^{i n}(t) \hat d^{i n \dagger}\left(t^{\prime}\right); \hat d^{i n \dagger}(t) \hat d^{i n}\left(t^{\prime}\right) \right\rangle= \left[ (\mathcal{F}+1) \delta\left(t-t^{\prime}\right); \mathcal{F} \delta\left(t-t^{\prime}\right)\right] $ and  $\left\langle \hat d^{i n}(t) \hat d^{i n}\left(t^{\prime}\right); \hat d^{i n \dagger}(t) \hat d^{i n \dagger}\left(t^{\prime}\right) \right\rangle=  \left[ \mathcal{S}\delta\left(t-t^{\prime}\right);\mathcal{S} \delta\left(t-t ^{\prime}\right) \right]$ where $\mathcal{F}=\sinh ^2 r$, $\mathcal{S}= 1/2 \sinh 2r$ and $r$ represent the squeezing parameter. The input correlations for the magnons are $\left\langle \hat O_{p}^{i n}(t) \hat O^{i n \dagger}_{p}\left(t^{\prime}\right); \langle \hat O_{p}^{i n \dagger}(t) \hat O_{p}^{i n}\left(t^{\prime}\right) \right\rangle =  \left[(\mathcal{N}_{O_{p}}(\omega_{ O_{p}})+1)\delta\left(t-t^{\prime}\right); \mathcal{N}_{O_{p}}(\omega_{ O_{p}}) \delta\left(t-t^{\prime}\right) \right]$, where $T = \frac{\hbar \omega_{O_p}}{k_B  \ln\bigg(\frac{1+\mathcal{N}_{O_{p}}}{\mathcal{N}_{O_{p}}}\bigg)}$, $\mathcal{N}_{O_{p}}$ represents the equilibrium mean thermal magnon numbers of the two magnon modes, and $k_B$ is the Boltzmann constant. In order to study the dynamics of the quantum fluctuations, we express the field operators as the sum of their steady-state values and the fluctuations around these steady-state values, such that   $\mathcal{\hat R}_{j}=\left\langle \mathcal{\hat R} _{j}\right\rangle + \delta \mathcal{\hat R}_{j}$ with  $\mathcal{\hat R}_{j}$ replaces the two operators $\hat d$ and $\hat O_{p}$, $\left\langle \mathcal{\hat R} _{j}\right\rangle $ are the mean values in the steady state, and $\delta \mathcal{\hat R}_{j}$ are the operators of fluctuation \cite{eleuch}. The fluctuations within the system can be characterized by the QLEs
\begin{equation}
		\begin{aligned}
			\delta \dot{\hat d} & =-\left(i \Delta_d+k_d \right) \delta \hat d-i \mathcal{G}_{O_1 d} \delta \hat O_1-i \mathcal{G}_{O_2 d} \delta  \hat O_2+ 2\lambda \delta \hat d^{\dagger} +\sqrt{2 k_d} \hat d^{i n}, \\
			\delta \dot {\hat O}_1 & =-\left(i \Delta_{O_1}+k_{O_1}\right) \delta \hat O_1-i \mathcal{G}_{O_1 d} \delta \hat d+ \sqrt{2 k_{O_1}} \hat O_1^{i n}, \\
			\delta \dot{\hat O}_2 & =-\left(i \Delta_{O_2}+k_{O_2}\right) \delta \hat O_2-i \mathcal{G}_{O_2 d} \delta \hat d+\sqrt{2 k_{O_2}} \hat O_2^{i n} .
		\end{aligned}
	\end{equation}
The quadratures of the cavity field and the two magnon modes are given by $ \delta \mathcal{\hat X}= (\delta \hat d +\delta \hat d ^{\dagger} )/ \sqrt{2} $, $ \delta \mathcal{\hat P}= i (\delta \hat d^{\dagger} -\delta \hat d ) / \sqrt{2}$, $ \delta \hat x _{p} = (\delta \hat O^{\dagger}_{p} +\delta \hat O_{p})/ \sqrt{2} $ and  $ \delta \hat y _{p} = i(\delta \hat O^{\dagger}_{p} -\delta \hat O_{p})/ \sqrt{2} $, and similarly for the input noise operators, $ \delta \mathcal{\hat X} ^{in}= (\delta \hat d ^{in} +\delta \hat d ^{{\dagger}^{in}} )/ \sqrt{2} $, $ \delta \mathcal{\hat P}^{in}=i (\delta d^{{in}{\dagger}} -\delta \hat d^{in}  ) / \sqrt{2}$, $ \delta \hat x^{in} _{p} = (\delta \hat O_p^{{in}{\dagger}} +\delta \hat O_p^{in})/ \sqrt{2} $ and  $ \delta \hat y^{in} _{p} = i(\delta \hat O^{{in}{\dagger}}_{p} -\delta \hat O^{in}_{p})/ \sqrt{2} $. The QLEs describing the quadrature fluctuations ($\delta \mathcal{\hat X}, \delta \mathcal{\hat P},\delta \hat x _{1}, \delta \hat y _{1}, \delta \hat x _{2}, \delta \hat y _{2}  $ ) can be expressed  as
	\begin{equation}
	\dot{\mathcal{L}(t)}= \mathcal{ U} \mathcal{L}(t)+ \mathcal{q}(t), 
	\end{equation}
where $\mathcal{L}(t)=\left[\delta \mathcal{\hat X} (t) , \delta \mathcal{\hat P} (t) , \delta \hat x_1(t), \delta \hat y_1(t), \delta \hat x_2(t), \delta \hat y_2(t)\right]^T$, $\mathcal{q}(t)=\left[\sqrt{2 k_d} \mathcal {\hat X}^{i n}, \sqrt{2 k_d} \mathcal{\hat P}^{i n}, \sqrt{2 k_{O_1}} \hat x_1^{i n}, \sqrt{2 k_{O_1}} \hat y_1^{i n}, \sqrt{2 k_{O_2}} \hat x_2^{i n}\right.$,
$\left.\sqrt{2 k_{O_2}} \hat y_2^{i n}\right]^T$ and 
$$
\mathcal{U}	=\left[\begin{array}{cccccc}
		-k_d+ 2\lambda & \Delta_d & 0 & \mathcal{G}_{O_1 d} & 0 & \mathcal{G}_{O_2 d} \\
		-\Delta_d & -k_d - 2 \lambda & -\mathcal{G}_{O_1 d} & 0 & -\mathcal{G}_{O_2 d} & 0 \\
		0 & \mathcal{G}_{O_1 d} & -k_{O_1} & \Delta_{O_1} & 0 & 0 \\
		-\mathcal{G}_{O_1 d} & 0 & -\Delta_{O_1} & -k_{O_1} & 0 & 0 \\
		0 & \mathcal{G}_{O_2 d} & 0 & 0 & -k_{O_2} & \Delta_{O_2} \\
		-\mathcal{G}_{O_2 d} & 0 & 0 & 0 & -\Delta_{O_2} & -k_{O_2}
	\end{array}\right].
	$$
The system is defined as a continuous variable (CV) three-mode Gaussian state, and its comprehensive description involves a $6 \times 6$ covariance matrix (CM) denoted as $\mathcal{V}$, defined as follows
\begin{equation}
		\mathcal{V}(t)=\frac{1}{2} \left\langle \mathcal{L}_{i}(t) \mathcal{L}_{j}(t')+\mathcal{L}_{j}(t') \mathcal{L}_{i}(t)\right\rangle, \quad  (i,j = 1,2....6) 
\end{equation}
the CM $\mathcal{V}$ in its steady state can be acquired through the solution of the Lyapunov equation
	\begin{equation}
		\mathcal{U} \mathcal{V}+ \mathcal{V} \mathcal{U}^{T}=-D
	\end{equation}
where $D$ is the diffusion matrix  defined as  $\left\langle \mathcal{q}_{i}(t) \mathcal{q}_{j}(t')+\mathcal{q}_{j}(t') \mathcal{q}_{i}(t)\right\rangle /2 = D_{ij}  \delta (t-t')  $ it can  be expressed as  $D=D_d \oplus D_O$, with $ D_d=\operatorname{diag}\left[\kappa_d \left(2 \sinh ^2 r+1+\sinh 2r\right),  \kappa_d \left(2 \sinh ^2 r+1-\sinh 2r\right)  \right]$ and $ D_ O=\operatorname{diag}\left[\kappa_{O_1}\left(2 \mathcal{N}_{O_1}+1\right),  \kappa_{O_1}\left(2\mathcal N_{O_1}+1\right), \kappa_{O_2}\left(2 \mathcal{N}_{O_2}+1\right), \kappa_{O_2}\left(2 \mathcal{N}_{O_2}+1\right)\right]$. After obtaining the CM $\mathcal{V}$ for the steady-state system through the aforementioned calculation, we can then explore the bipartite and tripartite entanglement properties that exist among the three modes.

\section{quantum correlations}

We assess the non-classical correlations within the bipartite subsystem consisting of magnons $O_1$  and $O_2$ by employing logarithmic negativity, Gaussian quantum steering, and the GIP. We quantify the tripartite magnon-photon-magnon entanglement using the minimum residual contangle. The global CM of the two magnon modes can be simplified into the following matrix
\begin{equation}
	\label{covar}
	\mathcal{V}_{12}=\left[\begin{array}{cc}
		x & z \\
		z^T & y
	\end{array}\right]\equiv\left[\begin{array}{cccc}
		a & 0 & c & 0	\\
		0 & b & 0 & d	\\
		c & 0 & a & 0	\\
		0 & d & 0 & b
	\end{array}\right],
\end{equation}
the covariance matrices $x=diag(a,b)$ and $y=diag(a,b)$, each with dimensions $2 \times 2$, representing individual modes. The $2 \times 2$ CM $z=diag(c,d)$ characterizes the correlation between the two magnons. In the framework of CV systems, the logarithmic negativity $E_{O_1O_2}$ written as \cite{en0,en1,en2}
\begin{equation}
	\label{34}
	E_{O_1O_2} = \max \left[ 0,- ln(2\varrho^{-})\right], 
\end{equation}
where $\varrho^{-}$ is the smallest symplectic eigenvalue measuring the entanglement between the two magnons modes, given by 	
\begin{equation}
	\varrho^{-}=\left( \frac{\Delta-\sqrt{\Delta^2-4 \operatorname{det} \mathcal{V}_{12}}}{2}\right)^{1/2},
	\label{ji} 
\end{equation} 
the parameter $\Delta$ is given by $\Delta= 2(ab-cd)$. The two magnon are separable if $\varrho^{-} > \frac{1}{2}$ (i.e., $E_{O_1O_2}=0$). The Gaussian quantum steering  refers to the asymmetry observed between two entangled observers. This property enables one party (Alice) to influence or "steer" the state of a remote party (Bob) by exploiting the entanglement they share \cite{steer}. This property can be used as a measure of the degree of steerability between the two magnons. There are three  modes of steerability. The first is "no-way steering," which occurs when $S^ {O_{1}\rightarrow O_{2}} = S^ {O_{2}\rightarrow O_{1}} = 0$, indicating that magnon $1$ cannot steer the magnon $2$, and vice versa. The second is "two-way steering", which occurs when $S^ {O_{1}\rightarrow O_{2}} = S^ {O_{2}\rightarrow O_{1}} \textgreater 0$, indicating that the magnon $1$ can steer the magnon $2$ and vice versa. The third is "one-way steering" if  only one Gaussian  $O_{1}$  $\rightarrow $ $O_{2}$ or Gaussian $O_{2}$  $\rightarrow$ $O_{1}$ is steerable. It is important to note that a non-separable state does not always constitute a steerable state, whereas a steerable state is always non-separable \cite{steer pr amaz}. We employ the CM as expressed in Eq. (\ref{covar}) and write the Gaussian measurement on magnon $1$  as \cite{steer}.
	\begin{equation}
		\label{base steer}
		S^{O_{1} \rightarrow O_{2}}:=\max \left[0,-\ln \left(\bar{\varPsi}^{O_{2}}\right)\right], 
	\end{equation}
	where  $\bar{\varPsi}^{O_{2}}=\left( \operatorname{det} \mathcal{A}^{O_{2}} \right) ^{1/2}$ is the  simplectic eigenvalues of the matrix $\mathcal{A}^{O_{2}}$ defined as  $\mathcal{A}^{O_{2}}=y-z^T x^{-1} z$, where $x$ , $y$, and $z$ are the $2 \times 2 $ matrices defined in equation (\ref{covar}). The quantification of Gaussian steering from $O_{1}$ to $O_{2}$ is defined by \cite{steer}
\begin{equation}
		S^{O_{1} \rightarrow O_{2}}:=\max \left[0,  \ln  \left(\frac{\operatorname{det}(x)}{4 \operatorname{det}(\mathcal{V}_{12})}\right)^{1/2}  \right].
\end{equation}
 In the context of quantum metrology the GIP serves as an additional tool for assessing QCs within a bipartite system \cite{37gip}. It plays a crucial role in comprehending how QCs contribute to enhancing the precision of parameters in quantum metrology protocols \cite{adiso2}. This is essentially due to its reliability and its ease of computability. The GIP is employed to quantify QCs  beyond entanglement. Its expression is as follows \cite{adiso2}
\begin{equation}
\mathcal{I}=\frac{c+\sqrt{c^2+h q}}{2 h}, 
\end{equation}
with $c=(\alpha+\gamma)(1+\alpha+\gamma-D)-D^2$, $h=(D-1)(1+2\alpha+2 \gamma+D)$, and  $q=(\alpha+D)(\alpha^{2}-D)+\gamma(2 \alpha+\gamma)(1+\alpha)$,
where $\alpha=4ab$, $\gamma=4cd$, and $D=16\operatorname{det}(\mathcal{V}_{12})$ are local symplectic invariants. The Gaussian interferometric power plays a pivotal role in assessing the non-classical aspects of the quantum nature of correlations that extend beyond entanglement within a bipartite state \cite{adiso2}.\\
  
To explore the tripartite entanglement within the system under consideration, we employ the residual contangle denoted as $ R_{min}$ \cite{residual1} as a quantitative measure. The quantification of tripartite entanglement is provided by determining the minimum residual contangle

\begin{equation}
R_{\min } \equiv \min \left[R^{\mathrm{d} \mid \mathrm{O1O2}}, R^{\mathrm{O1} \mid \mathrm{O2d}}, R^{\mathrm{O2} \mid \mathrm{dO1}}\right] ,
\end{equation}
with $R^{ l \mid  m  n} \equiv C_{  l \mid  m n  }-C_{ l \mid  m}- C_{ {l} \mid  n} \geq 0 \quad   (l, m, n= d,  O1,  O2)$ represente the residual contangle and $C_{  f\mid  e}$  is the contangle of subsystems of $f$ and $e$, where $e$ contains either one or two modes. The contangle $C_{  f\mid  e}=E_{  f\mid  e}^{2} $. The presence of a nonzero minimum residual contangle implies that $R_{\min }>0$, indicating the existence of genuine tripartite entanglement within the system. $C_{l\mid mn}$ is the one-vs-two-mode entanglement between mode $l$ and modes $m+n$. When $C_{l\mid mn}>0$, i.e. all one-vs-two-mode bi-partitions in the system are inseparable, the tripartite negativity $R_{\min }>0$ implies the existence of genuine tripartite entanglement shared within the system \cite{GGiedke01}. To quantify $C_{l\mid mn}$, we employ the logarithmic negativity \cite{VidalPlenio}, which is calculated as 
\begin{equation}
C_{l \mid m n}=\max \left[0, ln (2 	\varrho^{-})\right],
\end{equation} 
with $\varrho^{-}= \text{min eig}\left| i\Omega_3(P_{l\mid mn}\mathcal{V} P_{l\mid mn})\right| $, where $\Omega_3=\oplus^{3}_{j=1} i \sigma_y$ and $P_{l\mid mn}$ is the matrix that inverts the sign of momentum of mode $l$.

\section{Results and discussions}

In order to contrast our findings with protocols employing nonlinear techniques, a recent investigation \cite{26dens} identified entanglement levels between magnon modes that approached $0.25$ at a temperature of $10$ $\mathrm{mK}$ by utilizing Kerr nonlinearity induced by a strong classical drive \cite{26dens}. Another study \cite{25dens} employed a different type of nonlinearity, specifically magnetostrictive interaction within a YIG sphere, and produced similar entanglement levels at the same temperature. However, the entanglement diminishes as the temperature approaches $20$ $\mathrm{mK}$. Other way, by just employing a weak squeezed vacuum field to drive the cavity without any non-linearities, they achieve the production of a robust and steady entanglement spanning temperatures ranging from $0$ to $100$ $\mathrm{mK}$ \cite{ lien}. Our proposed method to generate entanglement produces a steady and strong entanglement between $0$ and $120$ $\mathrm{mK}$, and a significant degree of entanglement is present even at $500$ $\mathrm{mK}$ by using the nonlinearity of the OPA and by driving the cavity with a squeezed vacuum field. We employ a collection of parameters that have been determined to be experimentally feasible \cite{16}: $ \omega_{d} / 2 \pi= 10$ $\mathrm{GHz}$, $k_d / 2 \pi=5 k_{O_i} / 2 \pi=5$ $ \mathrm{MHz}$,  $\mathcal{G}_{O d}=\mathcal{G} _{O_1 d} = \mathcal{G}_{O_2 d}=4 k _d$ and $T=20$ $\mathrm{mK}$, we consider that $\mathcal{N}_{O_{1}}= \mathcal{N}_{O_{2}}\approx 0$ at $20 \mathrm{mK}$. The diameter of the YIG sphere is  250-$\mu m$, and the number of spins $n \approx 3.5 \times 10^{16}$. We've set the parameters to make the two magnon modes identical. We note that $\Delta _{O}=\Delta _{O_1} =  \Delta _{O_2}$, in other terms, $ \omega_{d} = \omega_{O_{1}}= \omega_{O_{2}} $ and that $ k_{O_1}=k_{O_2}$.
\begin{figure}[h]
	\centering
	\includegraphics[width=1\linewidth]{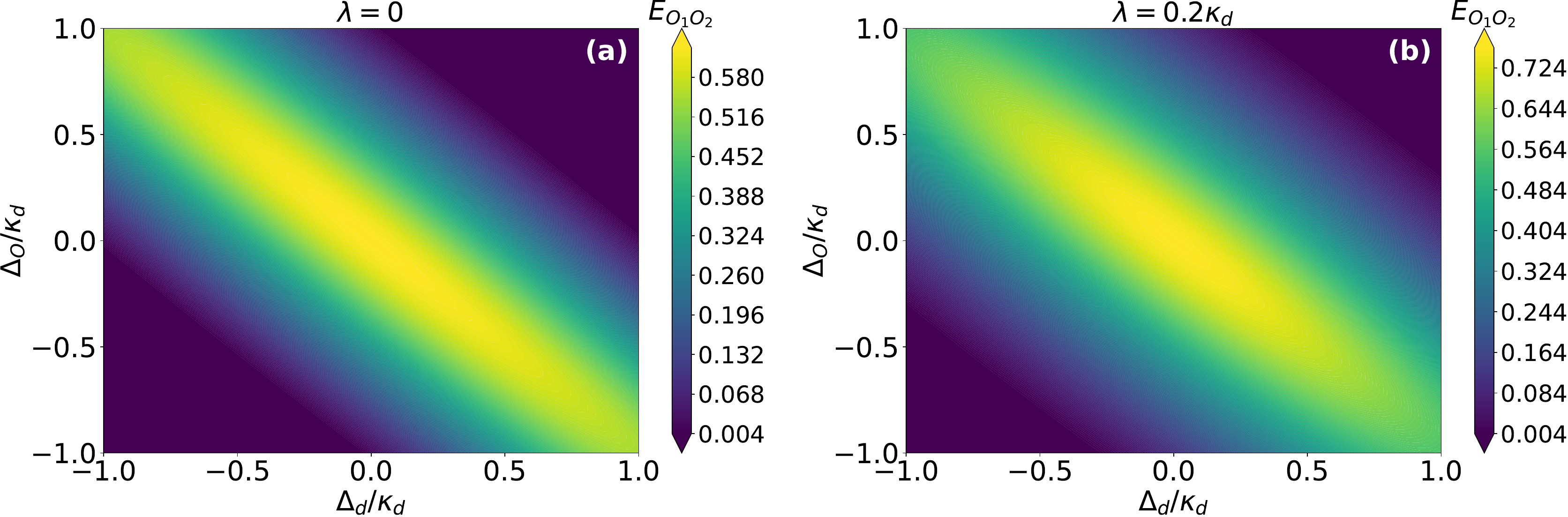}
	\caption{Density plot of bipartite entanglement  $E_{O_{1}O_{2}}$ between the two magnon modes versus the detunings $\Delta_{O}$ and $\Delta_{d}$ for $\lambda=0$ (a), for $\lambda= 0.2 k_d$ (b). The  other parameters are indicated  in the text.}
	\label{C}
\end{figure}

We start in Fig. \ref{C} by exploring the variation of the bipartite entanglement between the two magnon modes $O_1$ and $O_2$ as a function of the detunings $\Delta_{O}$ and $\Delta_{d}$ for different values of the gain $ \lambda$. We remark that the entanglement between the two magnons is optimal when $\Delta_{O}=\Delta_{d}=0$, i.e., $\omega_{d}=\omega_{u}$ and $\omega_{O}=\omega_{u}$. The entanglement increases with the increase in the gain of the OPA. In the case of $\lambda = 0$ (absence of the OPA), the maximum value of entanglement is almost equal to $0.580$ \cite{26dens}, and for $\lambda = 0.2k_d$, the maximum amount of entanglement is almost equal to $0.724$. Thus, the entanglement between the two magnons is significantly enhanced by the presence of the OPA.

	 \begin{figure}[h]	 
	\includegraphics[scale=0.41]{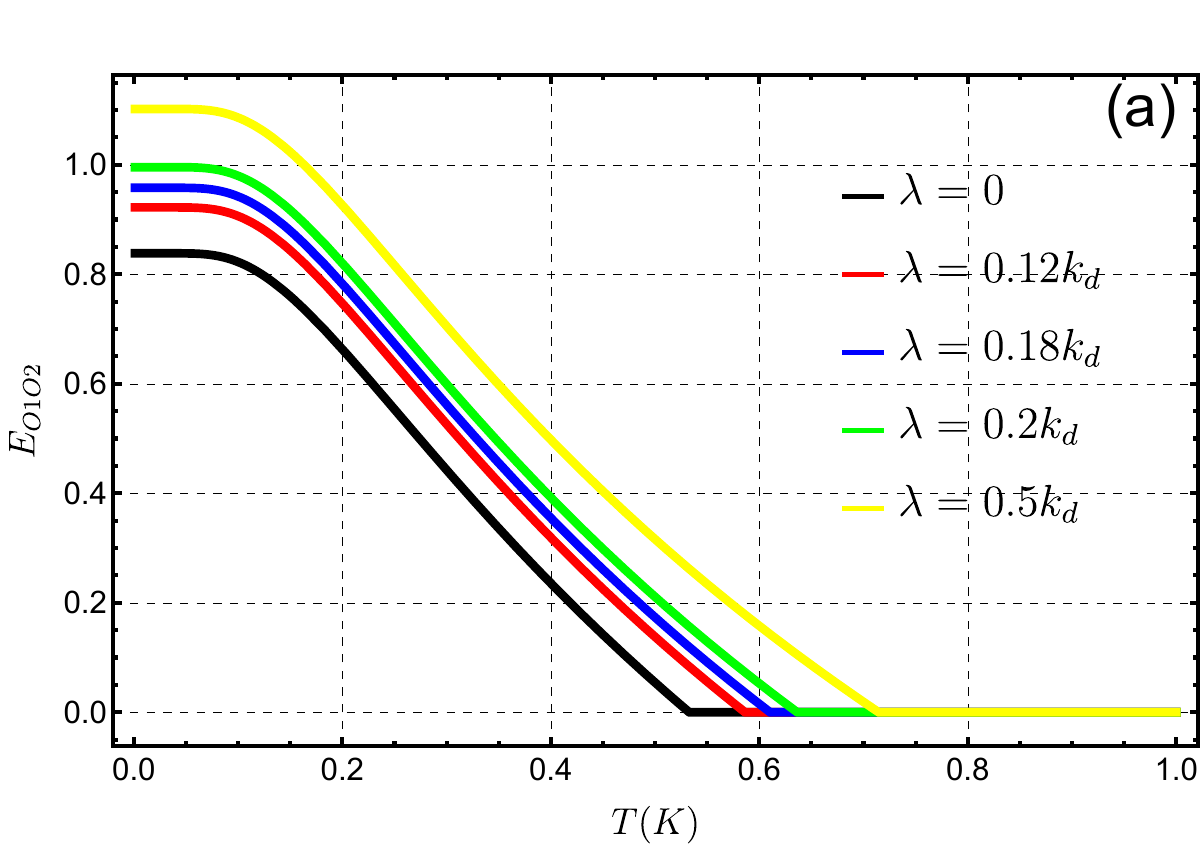} 
	\hfil
	\includegraphics[scale=0.41]{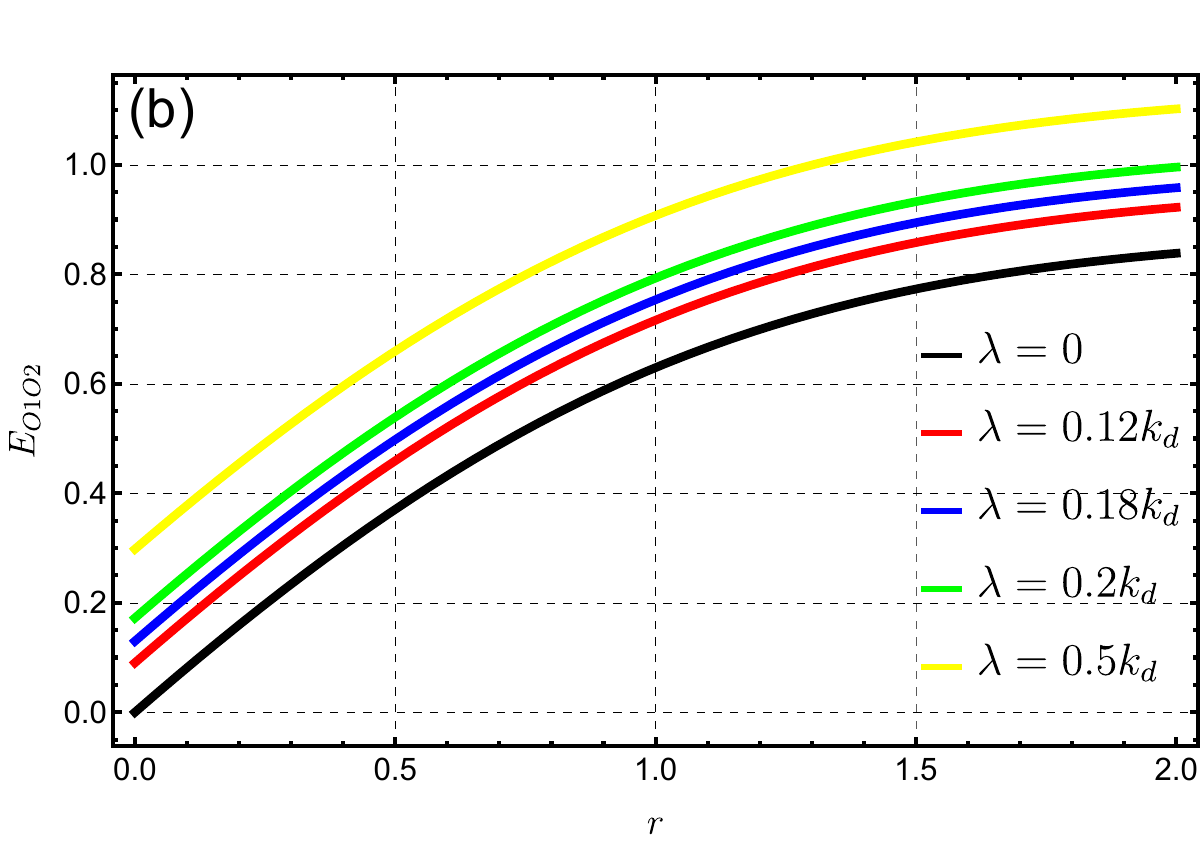}
	\caption{(a) Plots of bipartite entanglement $E_{O_{1}O_{2}}$ between the magnons modes against temperature $T$ for various values of the gain $\lambda$ with $r=2$. (b) Plots of bipartite entanglement $E_{O_{1}O_{2}}$ between the magnons modes versus the squeeze  parameter $r$ for various values of the gain $\lambda$ at $T=20 \mathrm{mK}$.}
		\label{A}
	\end{figure}	

Next, in Fig. \ref{A}(a), we illustrate the variation of logarithmic negativity $E_{O_{1}O_{2}}$ versus the temperature for different gain values. We note that for a given value of $\lambda$, the logarithmic negativity decreases with increasing $T$ and vanishes around $T\textgreater 0.7$ $\mathrm{K}$  for $\lambda= 0.5 k_{d}$. This decrease in entanglement as a function of temperature can be explained by the phenomenon of decoherence \cite{58 mir mir}. As the gain $\lambda$ increases, the temperature at which entanglement disappears increases. Consequently, entanglement exhibits increased resilience against temperature and becomes more robust with higher gain values. We also note an enhancement in the entanglement as the nonlinear gain $\lambda$ increases. The maximum entanglement value observed without the OPA present ($\lambda=0$) is approximately $E_{O1O2}$ $\simeq 0.85$ \cite{26dens}. However, with the introduction of the OPA, the maximum value of entanglement reaches  $E_{O1O2}$ $\simeq 1.1$ for  $\lambda= 0.5 k_d$. This highlights the increased entanglement between the two magnon modes when employing the OPA, in comparison to the outcomes documented in the study conducted in the absence of the OPA \cite{lien}.
 Then, in Fig. \ref{A}(b), we present the logarithmic negativity $E_{O_{1}O_{2}}$ versus the squeezing parameter $r$ for different values of $\lambda$. We note that entanglement increases rapidly as the squeezing parameter $r$ increases. Also, we observe that entanglement increases with increasing nonlinear gain $\lambda$ of the OPA. This illustrates that the magnon-magnon entanglement is deeply dependent on the squeezing parameter $r$, as depicted in Ref \cite{onewayster, am9}.


	\begin{figure}[h]
	\includegraphics[scale=0.41]{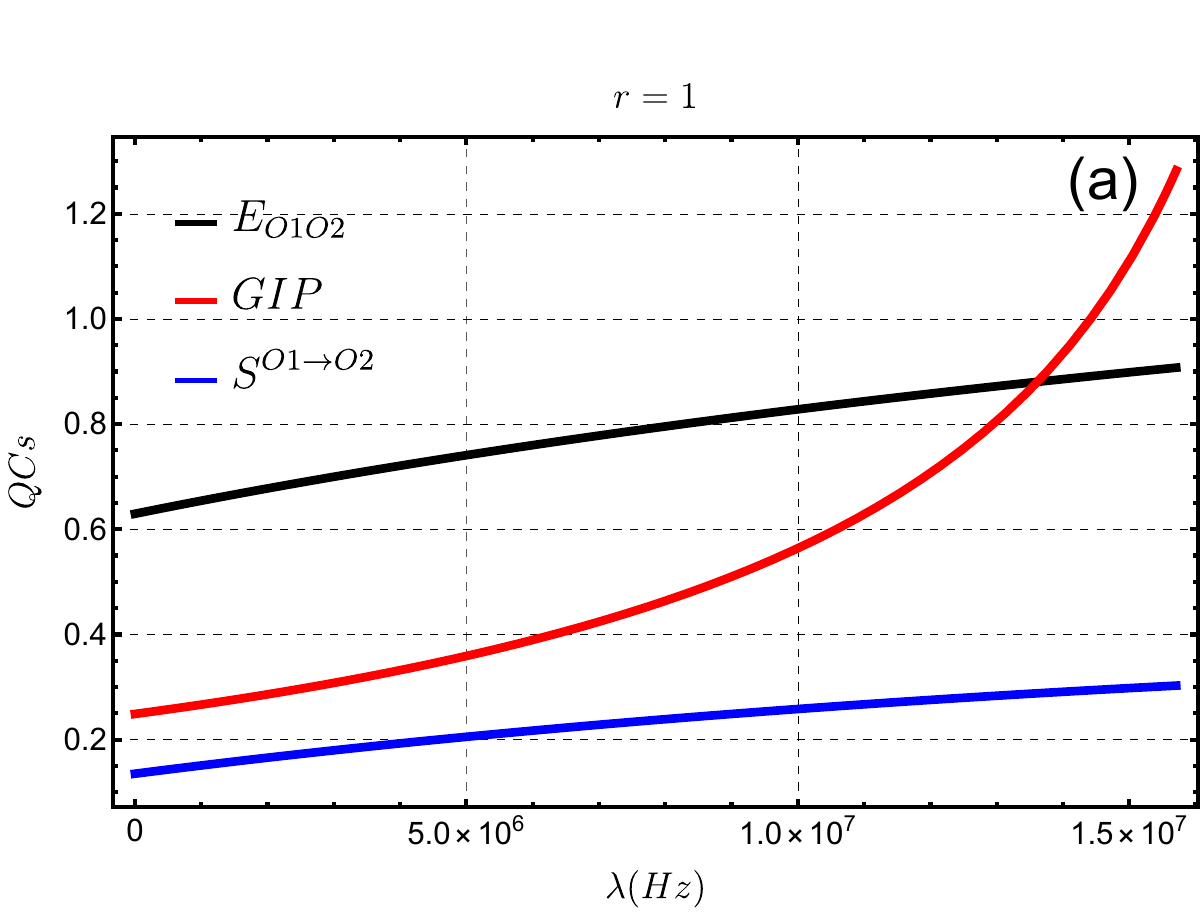}
	\hfil
	\includegraphics[scale=0.41]{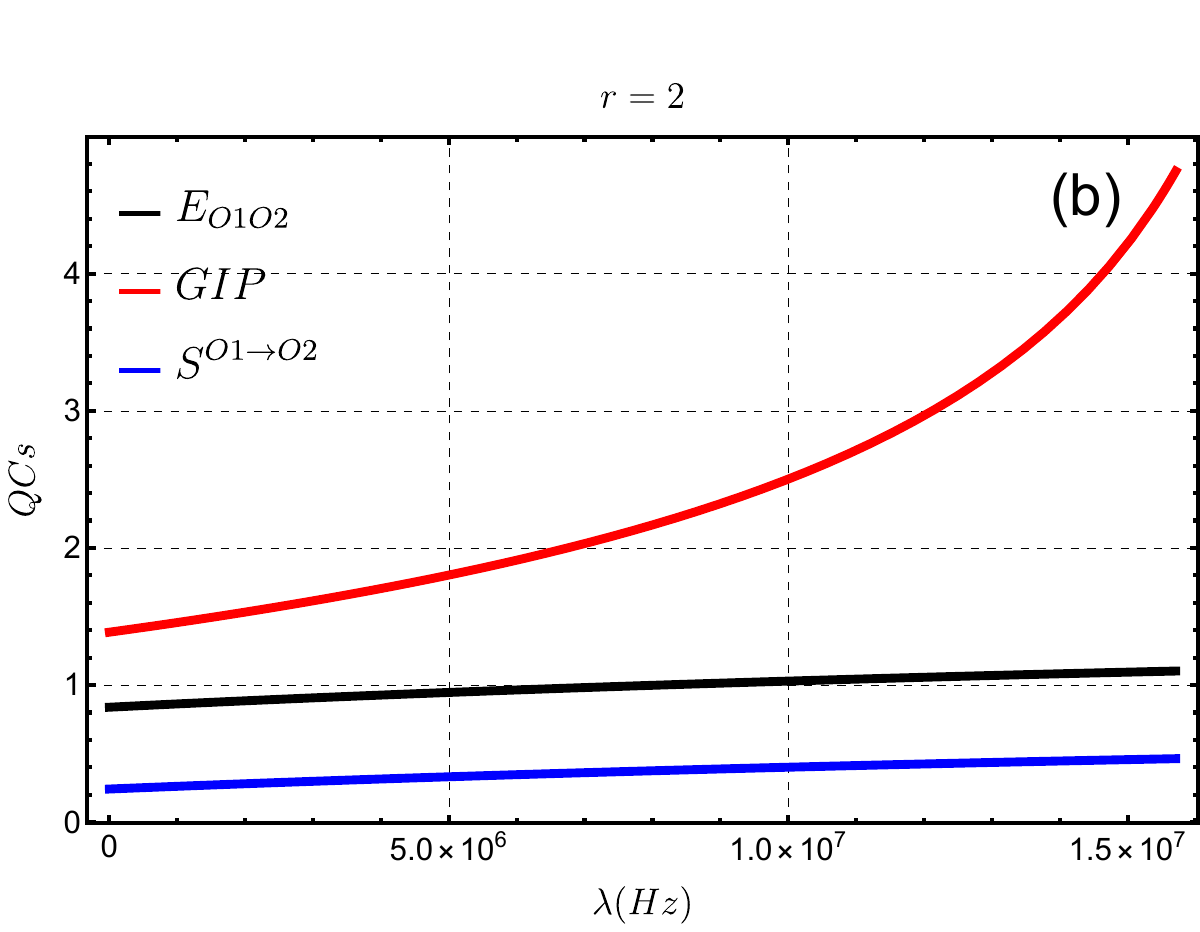}
	\caption{Plots of the QCs, e.g., steering $S^{O_{1}\rightarrow O_{2}}$, logarithmic negativity $E_{O_{1}O_{2}}$ and the GIP of the two magnon modes versus the gain $\lambda $ of the optical parametric amplifier for $ r=1$ (a) and for $r=2$ (b). $S^{O_{2}\rightarrow O_{1}}= S^{O_{1}\rightarrow O_{2}} $.}  
	\label{B}
\end{figure}
In addition, Fig. \ref{B} shows the behavior of entanglement, GIP, and  Gaussian quantum steering $S^{O_{1}\rightarrow O_{2}}$ as a function of the gain of the parametric amplifier $\lambda$ for different values of the squeezing parameter $r$. The plots show that QCs (entanglement, GIP, and Gaussian quantum steering) increase with the increase of the gain $\lambda$.  The GIP monotonic increases with the increase of $\lambda$; we additionally note an augmentation in QCs between the two magnons with an increase in the parameter $r$, underscoring a robust reliance between them and the squeezing parameter $r$. The two magnon modes are two-way steering ($S^{O1\rightarrow O2} = S^{O2\rightarrow O1}> 0$ ), i.e., magnon $O_{1}$ can steer the magnon $O_{2}$ and  vice versa. We show that the amount of entanglement is more important than steering, but GIP is more robust than entanglement. The GIP is significantly enhanced with the presence of OPA and squeezed light.

We discuss the criteria for entanglement in a two mode CV system. We define a  novel set of operators $\mathcal{B}=(\hat{O}_{1}+\hat{O}_{2})/ \sqrt{2}$ and $\mathcal{C}= (\hat{O}_{1}-\hat{O}_{2})/ \sqrt{2}$. The Mancini criteria propose that for entangled two modes, they must satisfy the   following inequality 
 \begin{equation}
   \left\langle \delta {\mathcal{B}_{x}}^{2}\right\rangle  \left\langle \delta {\mathcal{C}_{y}}^{2}\right\rangle \leq 1/4
 	\label{lol}
 	\end{equation}
we consider that  $ \Delta_{12}= \left\langle \delta {\mathcal{B}_{x}}^{2}\right\rangle  \left\langle \delta {\mathcal{C}_{y}}^{2}\right\rangle$, $ \delta {\mathcal{B}_{x}}$ and $ \delta {\mathcal{C}_{y}}$ are the fluctuations in the quadratures $\mathcal{B}_{x}$ and $\mathcal{C}_{y}$  defined as  $\mathcal{B}_x=\left(\mathcal{B} +\mathcal{B}^{\dagger}\right) / \sqrt{2}, \mathcal{C}_{y}=i\left(\mathcal{C}^{\dagger}-\mathcal{C} \right) / \sqrt{2}$. The violation of the Eq. (\ref{lol}) implies that the YIG samples are separable. 
 \begin{figure}[h]	   
 	\includegraphics[scale=0.45]{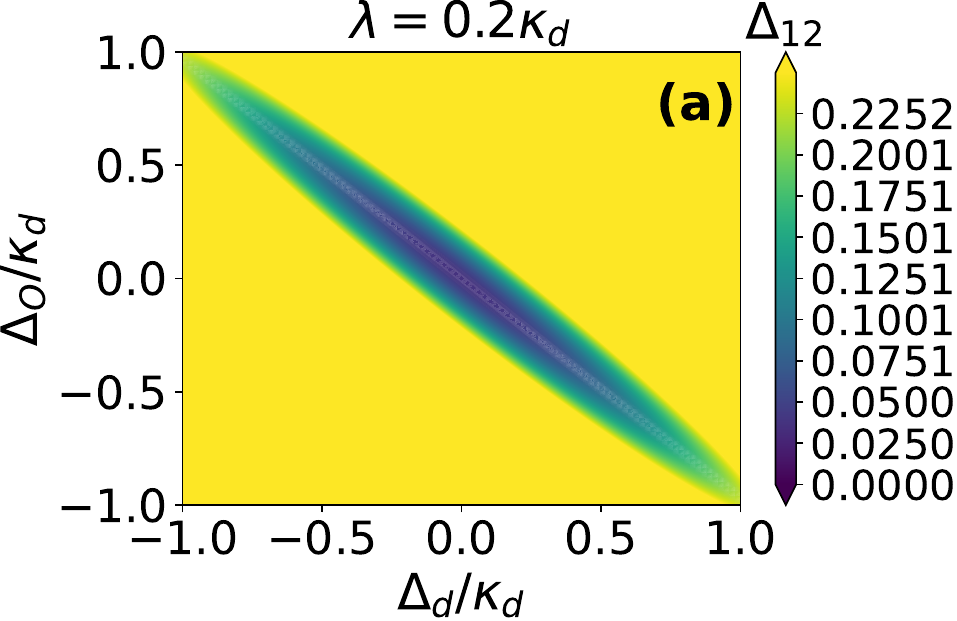}
 	\hfil	
 	\includegraphics[scale=0.45]{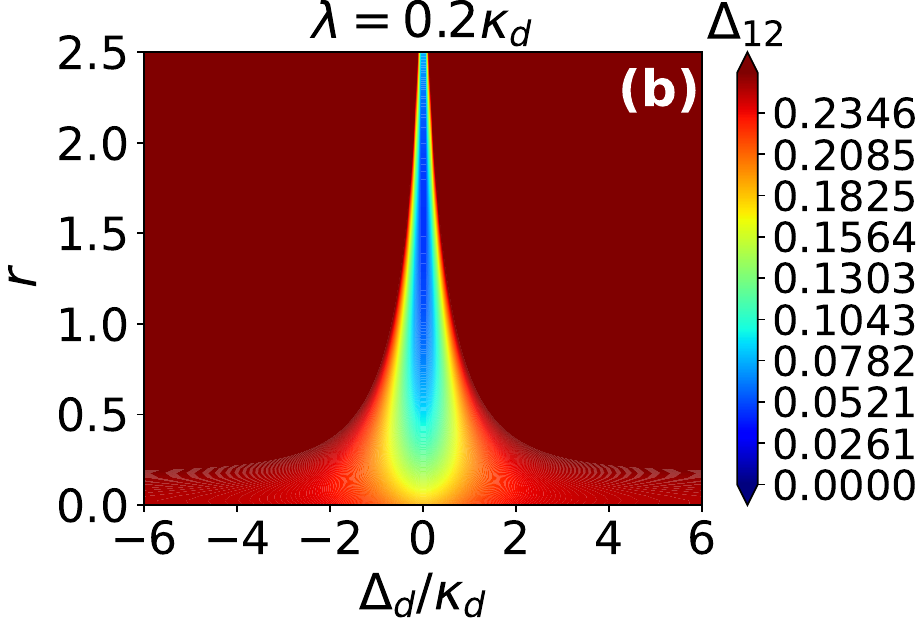}
 	\caption{(a) Plot of $\Delta_{12}$ as a function of the detuning $\Delta_{O}$ and $\Delta_{d}$ with $T=20$mK, $r=2$ and $\lambda=0.2k_d$. (b) Plot of $\Delta_{12}$ as a function of the detuning $\Delta_{d}$ and the squeezing parameter $r$ with $T=20$mK, $\Delta_O=0$ and $\lambda=0.2k_d$.}
 	\label{e}
 \end{figure}
 
 We plot in Fig. \ref{e}(a) the product $\Delta_{12}$ as a function of the detuning $\Delta_{O}/k_d$ and $\Delta_{d}/k_d$, where the product $\Delta_{12}$ is less than $1/4$. The figure shows that the two magnons are entangled when the cavity is fed by squeezed light and contains inside OPA. We remark that the entanglement of the two magnons is optimal around $\Delta_{d}=\Delta_{O}=0$. Moreover, in Fig. \ref{e}(b) we plot the $\Delta_{12}$ as a function of  $r$ and the detuning   $\Delta_{d}/k_d$ where the product $\Delta_{12}$ is less than $1/4$. Notice that the entanglement of the two magnons is optimal around $\Delta_{d}=0$. This entanglement is enhanced by squeezing parameter $r$ around $\Delta_{d}=0$, as depicted in Fig. \ref{e}(b). This is compatible with the numerical results in Fig. \ref{A}(b) about the variation of the logarithmic negativity against the squeezing parameter $r$; they both show that entanglement increases with the increase in the parameter of the squeezed field. 

\begin{figure}[h]	
 	\includegraphics[scale=0.45]{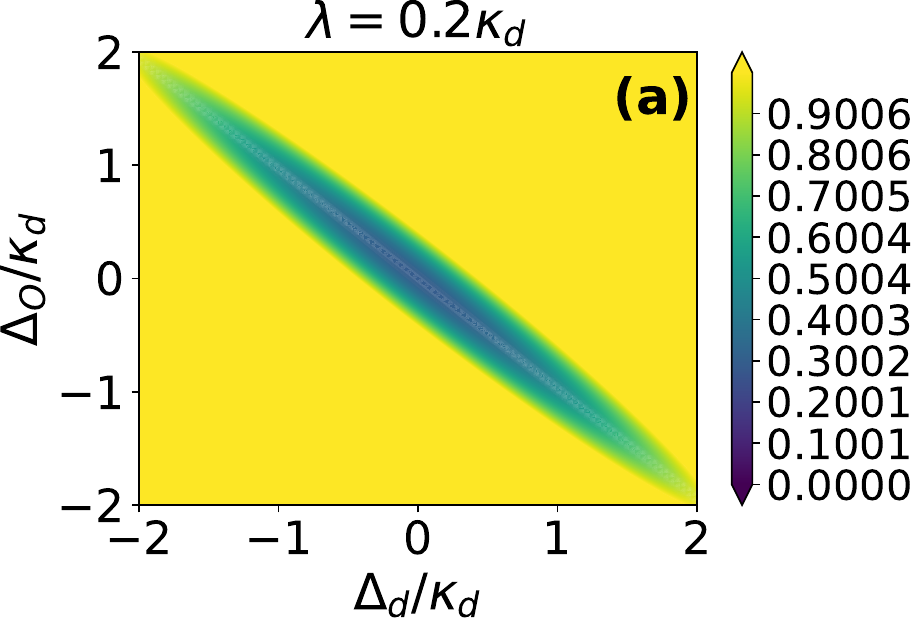} 
 	\hfil	 
 	\includegraphics[scale=0.45]{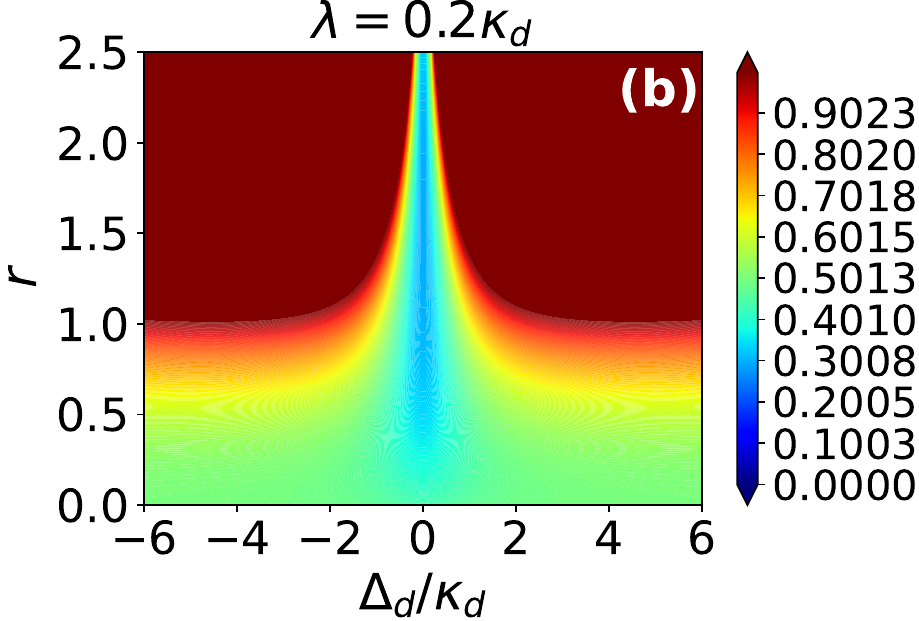}
\caption{(a) The variance of the quadrature fluctuation $\langle\delta \hat x_{1}(t)^2\rangle$ as a function of the detuning $\Delta_{O}$ and $\Delta_{d}$ with $T=20$mK, $r=2$ and $\lambda=0.2k_d$. (b) The variance of the quadrature fluctuation $\langle\delta \hat x_{1}(t)^2\rangle$ as a function of the squeezed parameter $r$ and the detuning $\Delta_{O}$ with $T=20$mK, $\Delta_O=0$ and $\lambda=0.2k_d$.}
\label{p}
\end{figure} 

As part of our research, we explore the squeezing of the one magnon mode and demonstrate that it can be achieved by using a squeezed vacuum field to drive the cavity and by utilizing interactivity squeezed light. The level of squeezing in a mode quadrature $\widehat Z$ can be represented in decibels (dB), derived from the  expression $-10 \log _{10}\left[\left\langle\delta \mathcal{\widehat Z}(t)^2\right\rangle /\left\langle\delta \mathcal{\widehat Z}(t)^2\right\rangle_{v a c}\right]$ with $\left\langle\delta \mathcal{\widehat Z}(t)^2\right\rangle_{v a c} = 1/2 $ \cite{lien}, 
is the variance of the vacuum state. The Fig. \ref{p} illustrates the region where we have squeezed state of  the  magnon 1, resulted from the squeezing vacuum noise of the cavity mode and the optomagnonics interaction. In Fig. \ref{p}(a), we observe that the variance of the quadrature fluctuation $\langle\delta \hat x_{1}(t)^2\rangle$ is optimal around $\Delta_{O}=\Delta_{d}=0$ (i.e, $\omega_{O}=\omega_{d}=\omega_u)$. The numerical results presented in Fig. \ref{p} (b), where we have depicted the steady-state variance of the quadrature fluctuation $\delta \hat x_{1}(t)$ as a function of the squeezed parameter $r$ and the detuning $\Delta_{O}$, illustrate that the suggested system allows for the generation of a squeezed state of magnons. We remark that the Figures \ref{p}(a) and \ref{p}(b) depict squeezing with a blue color. We noted the input squeezing is approximately 7dB for the magnon modes $O_{1}$ with $r\approx 2$. 


\begin{figure}[h!]
	\hfil
	\includegraphics[scale=0.41]{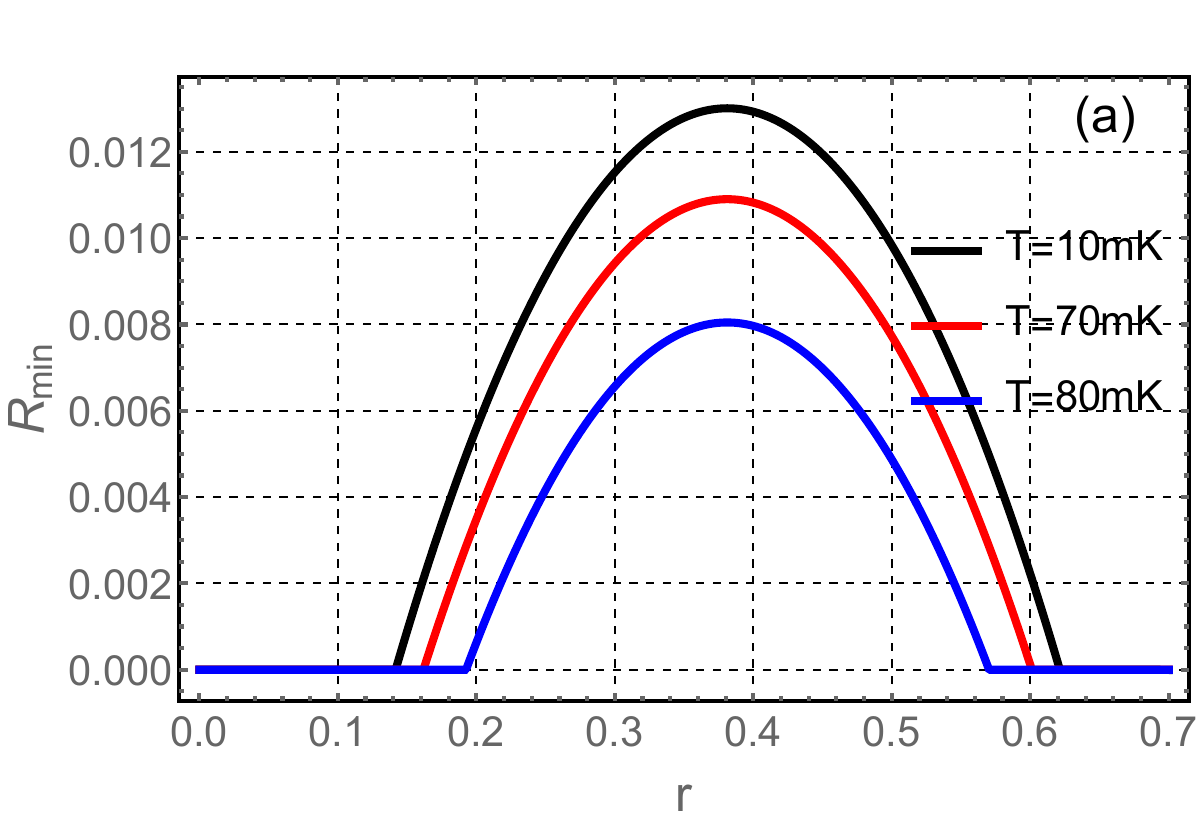}
	\includegraphics[scale=0.41]{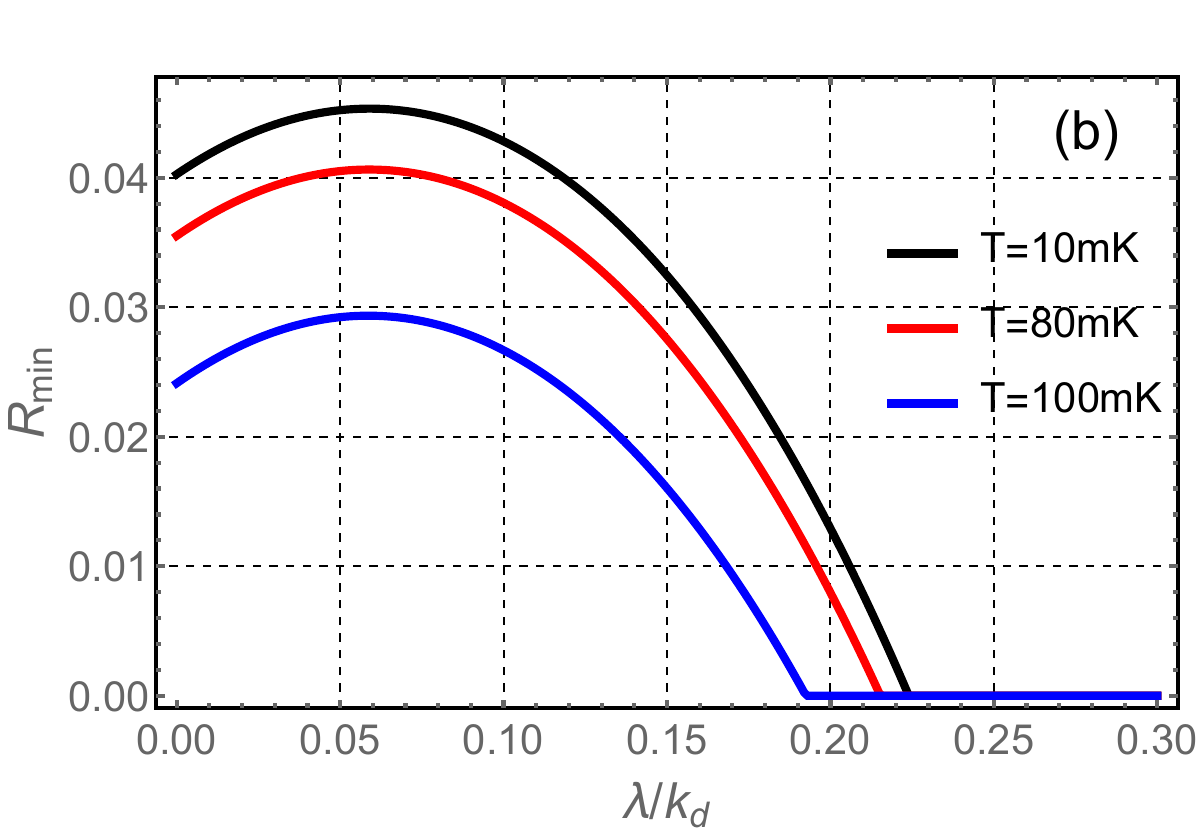}
\caption{ (a) Plot of tripartite entanglement in terms of the minimum residual contangle $R_{min}$ versus  the parameter $r$ of squeezed light for various values of the temperature $T$, with $\lambda=0.2k_d$ and $\mathcal{G}_{O d}=1 k _d$. (b) Plot of tripartite entanglement in terms of the minimum residual contangle $R_{min}$ versus the gain $\lambda$  for various values of the temperature $T$, with $r=0.4$.}
	\label{PO}
\end{figure}
The Fig. \ref{PO}(a) shows the variation of tripartite entanglement in terms of the minimum residual contangle $R_{min}$ as a function of the squeezing parameter $r$ for various values of the environmental temperature $T$. The figure shows that achieving entanglement among the three modes necessitates reaching a minimum value $r_{min}$ of the parameter $r$, which is greater than zero ($r_{min} > 0$). For a fixed value of the temperature $T$, the entanglement is achieved when $r$ exceeds $r_{min}$; the $r_{min}$ grows with increasing values of $T$, a phenomenon known as entanglement sudden birth \cite{1234,steer pr amaz}. Additionally, we see that for a given value of $T$, the quantity of entanglement grows as $r$ increases until it reaches its maximum value. However, after reaching the maximum value, the entanglement diminishes rapidly. This drop, even while $r$ increases, can be attributed to the reducing influence of radiation pressure \cite{am9}.\\

Finally, Fig. \ref{PO}(b) depicts the minimum residual contangle $R_{min}$ as a function of the nonlinear gain $\lambda /k_d$ for various values of the temperature $T$. It can be seen that the nonlinear gain  $\lambda/k_d$ contributes significantly to the emergence of genuine tripartite entanglement in the steady-state regime. For a fixed value of $T$, the degree of tripartite entanglement increases when the nonlinear gain increases until it reaches its maximum value, then decreases until it disappears, despite the increase in $\lambda/k_d$. For $T=10$ mK, the tripartite entanglement  vanishes when the   non-linear gain $\lambda/k_d$ exceeds approximately $0.225$. The amount of tripartite entanglement degrades with the increase in temperature under the effect of decoherence; for example, the level of $R_{min}$ reaches $0.046$ for $T = 10$ mK, unless for $T = 100$ mK, the maximum value reached by $R_{min}$ doesn't exceed $0.03$.\\

We conclude by exploring how to verify the generated entanglement, whether bipartite or tripartite. This verification can be achieved by measuring the corresponding CMs as described in \cite{am4,Teufel13}. The homodyne detection technique enables the straightforward measurement of the cavity field's quadratures by examining the cavity output. A weak microwave probe field is used to indirectly infer the magnon state. This involves homodyning the cavity output of the probe field. Crucially, the magnon mode's dissipation rate needs to be significantly smaller than the cavity mode (we have used $k_d=5k_O$). This ensures that when the driving field is switched off and the cavity photons decay, the magnon state remains practically unchanged. Only then is a weak probe field introduced to readout the preserved magnon information.

\section{Summary and Conclusions}
In summary, we have proposed a scheme to enhance QCs between non-interacting two magnon modes and tripartite magnon-photon-magnon entanglement in an  cavity magnonic  system through both an optical parametric amplifier and squeezed vacuum injection. We find that increasing the non-linear gain of OPA enhances the magnon-magnon entanglement against variations in environmental temperature. Moreover, manipulation of steady-state bipartite magnon-magnon entanglement is achievable by adjusting the squeezing parameter r. Furthermore, an increase in the nonlinear gain $\lambda$ and the squeeze parameter $r$ can improve QCs. The GIP behaves robustly in comparison to Gaussian suzntum steering and entanglement. The GIP is significantly enhanced with the presence of OPA and squeezed light. The presence of shared entanglement between the two samples of magnon is proved by using the Mancini criterion. The interaction between the cavity field and the magnon leads to the squeezed state of the first magnon. A zero value of the detunings is an optimal value for the squeezed state of the magnon mode. Additionally, within experimentally feasible parameters, a genuine magnon-photon-magnon tripartite entanglement state is realized. Our results underscore the significance of the squeezing parameter and non-linear gain for enhancing magnon-photon-magnon tripartite entanglement. We hope that current experimental technology will allow the proposed scheme to be implemented.


\begin{thebibliography}{99}
\bibitem{yig}
M. Xiong, M. Wang, G. Q. Zhang and J. Chen. Physical Review A. 107, 033516 (2023).
\bibitem{2}
X. Zhang, C. L. Zou, L. Jiang and H. X. Tang. Physical Review Letters. 113, 156401 (2014).
\bibitem{4}
Y. Tabuchi, S. Ishino, T. Ishikawa, R. Yamazaki, K. Usami
and Y. Nakamura. Physical Review Letters. 113, 083603 (2014).
\bibitem{5}
 H. Huebl, C. W. Zollitsch, J. Lotze, F. Hocke, M. Greifenstein,
A. Marx, R. Gross and S. T. B. Goennenwein.  Physical Review Letters. 111, 127003 (2013).
\bibitem{6}
 M. Goryachev et al. Physical Review Applied. 2, 054002 (2014).
\bibitem{7}
L. Bai, M. Harder, Y. P. Chen, X. Fan, J. Q. Xiao and C. M. Hu. Physical Review Letters. 114, 227201 (2015).
\bibitem{8}		
D. Zhang, X. M. Wang, T. F. Li, X. Q. Luo, W. Wu, F. Nori
and J. Q. You. Npj Quantum
Information. 1, 1-6 (2015).
\bibitem{1kiteel}
C. Kittel.  Physical Review. 73, 155 (1948).
\bibitem{9}		
D. L. Quirion, Y. Tabuchi, A. Gloppe, K. Usami, and Y. Nakamura. Applied Physics Express. 12, 070101 (2019).
\bibitem{10}
Y. P. Wang, G. Q. Zhang, D. Zhang, T. F. Li, C.M. Hu and J. Q. You. Physical Review Letters. 120, 057202 (2018).
\bibitem{12}
M. Harder, Y. Yang, B. M. Yao, C. H. Yu, J. W. Rao, Y. S. Gui,
R. L. Stamps and C. M. Hu. Physical Review Letters. 121, 137203 (2018).
\bibitem{11}
 L. Bai, M. Harder, P. Hyde, Z. Zhang, C. M. Hu, Y. P. Chen and J. Q. Xiao.  Physical Review Letters. 118, 217201 (2017).
\bibitem{13}
 X. Zhang, C. L. Zou, N. Zhu, F. Marquardt, L. Jiang and H. X. Tang. Nature communications. 6, 8914 (2015).
\bibitem{14}
D. Zhang, X. Q. Luo, Y. P. Wang, T. F. Li and J. Q. You. Nature communications. 8, 1368 (2017).
\bibitem{15}
 Y. Tabuchi, S. Ishino, A. Noguchi, T. Ishikawa, R. Yamazaki,
K. Usami and Y. Nakamura. Science. 349, 405 (2015).
\bibitem{16}
 X. Zhang, C. L. Zou, L. Jiang and H. X. Tang. Science advances. 2, e1501286 (2016).
\bibitem{17}
 B. Wang, Z. X. Liu, C. Kong, H. Xiong and Y. Wu. Optics Express. 26,  20248-20257 (2018).
\bibitem{18}
 C. Kong, B. Wang, Z. X. Liu, H. Xiong and Y. Wu. Optics Express. 27, 5544-5556 (2019).
 \bibitem{annal}
 M. Amazioug,  D. Dutykh, B. Teklu and  M. Asjad. Annalen der Physik. 536, 2300357 (2024).
  \bibitem{lien}
 	J. M. P. Nair and  G. S. Agarwal. Applied Physics Letters. 117, (2020).
\bibitem{Ann}
	X. Y. Yin, Z. B. Yang,  Y. M.  Huang, Q. M. Wan, R. C. Yang and H. Y.  Liu. Annalen der Physik. 535, 2200603 (2023).
	\bibitem{22}
	R. Reidinger, A. Wallucks, I. Marinković, C. Löschnauer, M. Aspelmeyer, S. Hong and S. Gröblacher. Nature. 556, 473-477 (2018).
	\bibitem{23}
	C. F. Ockeloen-Korppi, E. Damskägg, J. M. Pirkkalainen, M. Asjad, A. A. Clerk, F. Massel, M. J. Wooley and M. A. Sillanpää. Nature. 556, 478-482 (2018).
	\bibitem{super1}
	L. Dicarlo, M. D. Reed, L. Sun, B. R. Johnson, J. M. Chow, J. M. Gambetta, L. Frunzio, S. M. Girvin, M. H. Devoret and  R. J. Schoelkopf. Nature. 467, 574-578 (2010).
	\bibitem{super}
	E. Flurin, N. Roch, F. Mallet, M. H. Devoret and  B. Huard.  Physical Review Letters. 109, 183901(2012).

\bibitem{24}
 T. A. Palomaki, J. D. Teufel, R. W. Simmonds and K. W. Lehnert. Science. 342, 710-713 (2013).
\bibitem{secondnon} Y. Zhou, S. Y. Xie, C. J. Zhu and  Y. P.  Yang. Physical Review B. 106, 224404 (2022).
\bibitem{20}
J. Li, S. Y. Zhu  and G. S. Agarwal. Physical Review Letters. 121, 203601 (2018).
\bibitem{25}
 J. Li and S. Y. Zhu. New Journal of Physics. 21, 085001 (2019).
 \bibitem{28}
 E. Ginossar, M. Mirrahimi, L. Frunzio, S. M. Girvin and R. J. Schoelkopf. Nature. 495, 205-209 (2013).
 \bibitem{29}
 K. Tara, G. S. Agarwal and S. Chaturvedi. Physical Review A. 47, 5024 (1993).
   \bibitem{onewayster} Z. B. Yang, X. D. Liu, X. Y. Yin,  Y. Ming, H. Y. Liu and R. C. Yang. Physical Review Applied. 15, 024042 (2021).  	
   \bibitem{25dens}
  	J. Li, S. Y. Zhu. New Journal of Physics. 21, 085001 (2019).
  
 \bibitem{squerole1}
 C. M. Caves, K. S. Thorne,  R.W. Drever, V. D. Sandberg and  M. Zimmermann. Reviews of Modern Physics. 52, 341 (1980).
 \bibitem{squerole2}
 A. Bassi,  K. Lochan, S. Satin, T. P. Singh and H. Ulbricht. Reviews of Modern Physics. 85, 471 (2013).
 \bibitem{squezim}
 S. L. Braunstein and P. Van Loock. Reviews of Modern Physics. 77, 513 (2005).
 \bibitem{fb}
 M. Amazioug, S. Singh, B. Teklu and M. Asjad. Entropy. 25, 1462 (2023). 
 
 \bibitem{sohail}
A. Sohail, R. Ahmed, J. X. Peng, A. Shahzad and S. K. Singh. JOSA B. 40, 1359-1366 (2023). 
 \bibitem{steer}
 I. Kogias, A. R. Lee, S. Ragy and G. Adesso. Physical Review Letters. 114, 060403 (2015).
 \bibitem{lifetimelongmagnon}
 Q. Zheng, W. Zhong, G. Cheng and A. Chen. Quantum Information Processing. 22, 140 (2023).
 \bibitem{1}
 T. Holstein and H. Primako. Physical Review. 58, 1098 (1940).
  \bibitem{noise correlations}
 C. W. Gardiner. Physical Review Letters. 56, 1917 (1986).
\bibitem{eleuch}
E. A. Sete, H. Eleuch and C. R. Ooi. JOSA B. 31, 2821-2828 (2014).  
 \bibitem{en0}
 M. B. Plenio. Physical Review Letters. 95, 090503 (2005).
 \bibitem{en1}
G. Vidal and R. F. Werner. Physical Review A. 65, 032314 (2002).
  \bibitem{en2}
 G. Adesso, A. Serafini and F. Illuminati. Physical Review A. 70, 022318 (2004).  
\bibitem{steer pr amaz}
M. Amazioug and M. Daoud. The European Physical Journal D.  75, 178 (2021).
\bibitem{37gip}	X. Wang, B. C. Sanders. Physical Review A. 65, 012303 (2001).
	
\bibitem{adiso2}
G. Adesso. Physical Review A. 90, 022321 (2014).
 \bibitem{residual1}
G. Adesso and  F. Illuminati. Journal of Physics A: Mathematical and Theoretical. 40, 7821 (2007).
\bibitem{GGiedke01}
G. Giedke, B. Kraus, M. Lewenstein and J. I. Cirac. Physical Review A. 64, 052303 (2001).
 
\bibitem{VidalPlenio}
G. Vidal and R. F. Werner. Physical Review A. 65, 032314 (2002).
 \bibitem{26dens}
Z. Zhang, M. O. Scully and G. S. Agarwal. Physical Review Research. 1, 023021 (2019).
\bibitem{58 mir mir}
W. H. Zurek. Reviews of Modern Physics. 75, 715 (2003).
 \bibitem{am9} 
M. Amazioug, B. Maroufi and M. Daoud. Quantum Information Processing. 19, 16 (2020).
\bibitem{1234}
Z. Ficek and R. Tanas. Physical Review A. 74, 024304 (2006).
\bibitem{am4}	
D. Vitali et al. Physical Review Letters. 98, 030405 (2007).
\bibitem{Teufel13} 
T. A. Palomaki, J. D. Teufel, R. W. Simmonds, and K. W. Lehnert. Science. 342, 710 (2013).

 

	\end{thebibliography}
	\end{document}